\documentclass{aa}
\usepackage{psfig, wrapfig}

\begin{document}

   \title{Observation of GRB 030131 with the INTEGRAL satellite\thanks{Based on observations with INTEGRAL, an ESA project with instruments and science data centre funded by ESA member states (especially
the PI countries: Denmark, France, Germany, Italy, Switzerland, Spain), Czech Republic and Poland, and with the participation of Russia
and the USA,
and on observations collected by the Gamma-Ray
Burst Collaboration at ESO (GRACE) at the European Sourthern Observatory,
Paranal, Chile (Programme 70.D-0523).} }

   \author{D. G\"{o}tz\inst{1,2}, S. Mereghetti\inst{1},
    K. Hurley\inst{3}, S. Deluit\inst{4}, M. Feroci\inst{5}, F. Frontera\inst{6,7}, A. Fruchter\inst{8},
    J. Gorosabel\inst{8,16}, D.H. Hartmann\inst{9}, J. Hjorth\inst{10}, R. Hudec\inst{11}, I.F. Mirabel\inst{12,15},
    E. Pian\inst{13}, G. Pizzichini\inst{7}, P. Ubertini\inst{5} \and C. Winkler\inst{14}}

   \offprints{D. G\"{o}tz, email: diego@mi.iasf.cnr.it}

   \institute{Istituto di Astrofisica Spaziale e Fisica Cosmica -- CNR,
              Sezione di Milano ``G.Occhialini'',
          Via Bassini 15, I-20133 Milano, Italy
         \and
             Dipartimento di Fisica, Universit\`{a} degli Studi di Milano Bicocca,
             P.zza della Scienza 3, I-20126 Milano, Italy
         \and
         UC Berkeley Space Sciences Laboratory, Berkeley CA 94720-7450, USA
          \and
         Integral Science Data Centre, Chemin d'\'{E}cogia 16, CH-1290 Versoix, Switzerland
         \and
         Istituto di Astrofisica Spaziale e Fisica Cosmica -- CNR, via Fosso del Cavaliere 100, I-00133 Roma, Italy
         \and
         Dipartimento di Fisica, Universit\`{a} di Ferrara, Via Paradiso 12, I-44100 Ferrara, Italy
         \and
         Istituto di Astrofisica Spaziale e Fisica Cosmica -- CNR, Sezione di Bologna,  via Piero Gobetti 101, I-40129 Bologna, Italy
         \and
         Space Telescope Science Institute, 3700 San Martin Drive, Baltimore, MD 21218, USA
         \and
         Department of Physics and Astronomy, Clemson University, Clemson, SC 29634-0978, USA
         \and
         Astronomical Observatory, University of Copenhagen, Juliane Maries Vej 30, 2100 Copenhagen, Denmark
         \and
         Astronomical Institute, Academy of Sciences of the Czech Republic, CZ-251 65
Ondrejov, Czech Republic
         \and
         Service d'Astrophysique, CEA/Saclay, Orme des Merisiers B\^{a}t. 709, F-91191 Gif-sur-Yvette, France
         \and
         Osservatorio Astronomico di Trieste, Via G.B. Tiepolo 11, I-34131 Trieste, Italy
         \and
         ESA-ESTEC, RSSD, Keplerlaan 1, 2201 AZ Nordwijk, The Netherlands
         \and
         Instituto de Astronomia y Fisica del Espacio / CONICET, cc67, suc 28. 1428 Buenos Aires, Argentina
         \and
         Instituto de Astrof\'{\i}sica de Andaluc\'{\i}a (IAA-CSIC), P.O. Box 03004, E--18080 Granada, Spain
         }


\abstract{
A long Gamma-Ray Burst (GRB) was detected with the instruments on
board the INTEGRAL satellite on January 31 2003.
Although most of the GRB, which lasted $\sim$150 seconds,
occurred during a satellite slew, the automatic software of
the INTEGRAL Burst Alert System
was able to detect it in near-real time.
Here we report the results obtained with the IBIS instrument,
which detected GRB 030131 in the 15 keV - 200 keV energy range,
and ESO/VLT observations of its optical transient. The burst
displays a complex time profile with numerous peaks. The peak spectrum can be
described by a single power law with photon index $\Gamma\simeq$1.7 and has
a flux of $\sim$2 photons cm$^{-2}$ s$^{-1}$ in the 20-200 keV energy band.
The high sensitivity
of IBIS has made it possible for the first time
to perform detailed time-resolved spectroscopy
of a GRB with a fluence of 7$\times$10$^{-6}$ erg cm$^{-2}$ (20-200 keV).
}

\authorrunning{D. G\"{o}tz et al.}

\maketitle

%

\section{Introduction}
Ever since their discovery (\cite{klebesadel}),
Gamma-Ray Bursts (GRBs) have been a puzzling mystery, mostly because
of their short durations and the apparent
lack of counterparts at other wavelengths. A breakthrough
in this field came thanks to the Italian-Dutch
satellite {\it Beppo}SAX, which had the capability to
localize the bursts' prompt emission  with a
precision of a few arcminutes within a few hours. This led to the discovery
of the afterglow emission at lower energies, initially in X-rays
(\cite{costa}) and subsequently at optical (\cite{vanpa})
and radio (\cite{frail}) wavelengths, which allowed 
the redshift of these objects to be measured, and firmly established
the cosmological nature of GRBs.

Due to the limited duration and the fading
character of the afterglow emission, the prompt distribution
of GRB coordinates to the scientific community is a high priority.
After the end of the {\it Beppo}SAX mission this task has been accomplished
mainly by HETE-2 (\cite{ricker}). INTEGRAL (\cite {winkler}), although not
specifically designed as a GRB-oriented mission, can contribute
to the rapid localization of the prompt emission of GRBs thanks
to the INTEGRAL Burst Alert System (IBAS; \cite{mere2}).
This software, running at the INTEGRAL Science Data Centre
(ISDC; \cite{courvoisier}), is able to detect and localize
GRBs with a precision of a few arcminutes
in a few seconds, and to distribute their coordinates in near real time over the Internet.

The high sensitivity of the INTEGRAL instruments also allows us to study in detail
the prompt $\gamma$-ray emission of GRBs. This is particularly interesting for
the faintest bursts, for which deep spectral studies were not possible
up to now. For example, with the CGRO/BATSE instrument, time resolved spectroscopy
was possible only for bursts with a fluence larger
than $\sim$4$\times$10$^{-5}$ ergs cm$^{-2}$ (\cite{preece}).

On January 31 2003 at 07:38:49 UTC a GRB was detected in the field of view
of the main instruments on board INTEGRAL: IBIS (\cite{ubertini})
and SPI (\cite{vedrenne}). Here we concentrate on the results obtained with IBIS, a coded mask imaging telescope
based on two detectors, ISGRI and PICsIT, operating in the
15 keV - 1 MeV and 170 keV - 10 MeV energy ranges, respectively.

\section{Detection and Localization}

GRB 030131 was discovered by IBAS (using IBIS/ISGRI data) on January 31 2003 at 07:39:10 UTC
($\sim$ 21 seconds after the beginning of the GRB, see below).
The on-line automatic imaging analysis  localized it to off-axis angles
Z=10.1$^{\circ}$, Y=--3.6$^{\circ}$, in the
partially coded field of view (only $\sim$23\%
of the detector was illuminated by the GRB).
The GRB coordinates  were not distributed automatically
by IBAS  because  most
of the burst occurred during a satellite slew
(IBAS is disabled during satellite slews).
In fact only the first $\sim$20 seconds of this $\sim$150 s long burst,
during which the satellite attitude was stable and well known,
were analyzed by IBAS, resulting in a low significance of the trigger.
An off-line interactive analysis confirmed the reality of the event
(\cite {borkowski}), but the reported error region radius was underestimated.
A correct localization with an error radius of 5$'$ (\cite{mereghetti}) was distributed only
three days later.

\begin{figure}
      \hspace{0cm}\psfig{figure=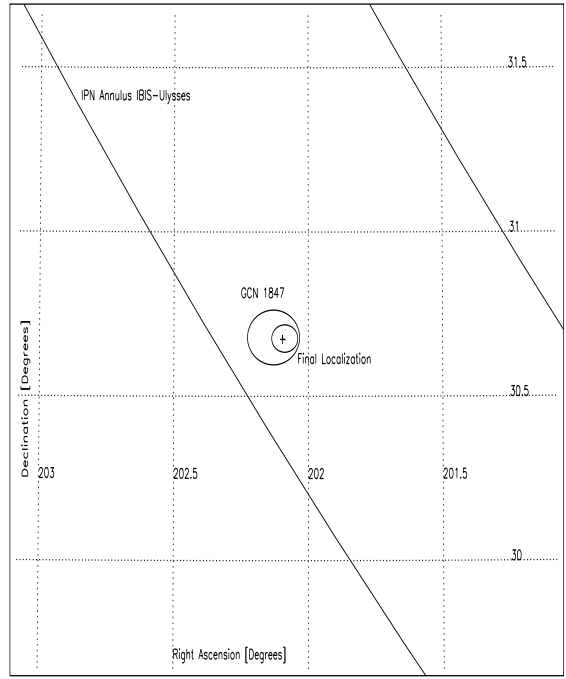,height=9.5cm,width=7.5cm,angle=90}
      \caption[]{Localizations of GRB 030131: the annulus obtained by the IPN
is consistent with the localization given in \cite{mereghetti} (GCN 1847)
and the one derived in this paper (the annulus obtained using SPI ACS and
Ulysses is consistent with the one plotted but has a larger width).
The cross indicates the position of the OT.}
         \label{ipn}
   \end{figure}

By accumulating data over short time intervals and analyzing the corresponding
images, we confirmed that the satellite slew started at
07:39:09 UTC, as indicated by the attitude data.
We therefore used the first 20 s of the event, corresponding to the stable
pointing period, to derive the GRB position
$\alpha_{J2000}$ = 13$^{h}$ 28$^{m}$ 21$^{s}$,  $\delta_{J2000}$ = +30$^{\circ}$
40$'$ 33$''$,
with an error radius of 2.5$'$.
Although the statistical error in these coordinates is only 1.6$'$,
we conservatively added a systematic uncertainty of 2$'$,
based on the results obtained in IBIS observations of sources
with known positions.
Our final position for GRB 030131 is consistent with  the one
reported  earlier (\cite {mereghetti})
and with the annulus derived with the IPN using Ulysses and IBIS/ISGRI data
(see Fig. \ref{ipn}).

A provisional identification of an
optical transient (OT) for GRB 030131 was reported by Fox et al. (2003a).
The candidate OT, detected
with the Palomar 48-inch Oschin telescope + NEAT Camera,
had  magnitude R$\sim$21.2 at 3.62 hours after the burst, but it
was much fainter and barely detectable (R$>$23.5)
$\sim$26.8 hours after the burst
at the 200-inch Hale telescope.
Its  coordinates,
$\alpha_{J2000}$ = 13$^{h}$ 28$^{m}$ 22.29$^{s}$,
$\delta_{J2000}$ = +30$^{\circ}$ 40$'$ 23.7$''$,
are 20$''$ from the center of the error circle derived here.

As a follow-up, we obtained a 3$\times$300 s exposure in the V band using the
European Southern Observatory Very Large Telescope (VLT) with the FORS1
instrument at a mean date of 13 February 09:11:54 UTC.
The seeing was about 1$''$.
There was no detectable object at the location of the candidate optical counterpart,
with a  5 $\sigma$ upper limit of  V$>$26.4.

The marginal detection $\sim$29 hours
after the burst,  with B = 25.4$\pm$0.3  (\cite{gorosabel}, but see also \cite{henden}),
and our VLT upper limit, confirm that this object is the OT of GRB 030131. Thus GRB 030131 is
the first GRB detected with INTEGRAL with an associated optical counterpart.

\section{Temporal and Spectral Analysis}

We have analyzed IBIS/ISGRI single events, for which arrival time, energy
deposit and interaction pixel of the detector are known for each event.
\begin{figure}
      \hspace{0cm}\psfig{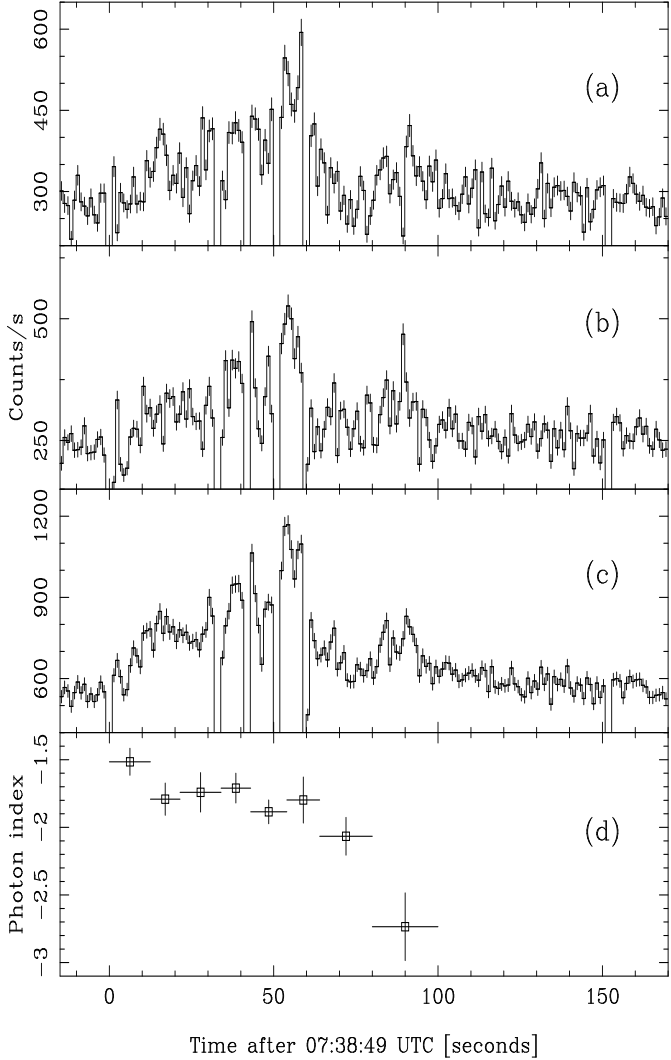}
      \caption[]{IBIS/ISGRI light curve of GRB 030131 in various energy bands
((a): 15-50 keV, (b): 50-300 keV, (c): 15-500 keV).
The six data gaps are artifacts caused
by satellite telemetry saturation. Four peaks can be identified,
after $\sim$15, $\sim$40, $\sim$55, $\sim$85 seconds. The light curve has been
corrected to take into  account the varying fraction  of the exposed instrument area
during the satellite slew. Panel (d) shows the spectral variation of the GRB with time.}
         \label{lc}
   \end{figure}

Fig. \ref{lc} shows the light curves of GRB 030131 binned at 1 s resolution
in different energy bands.
The burst started at 07:38:49 UTC and  lasted for about 150 s.
The time profile shows several peaks (note that the small gaps are
artifacts caused by satellite telemetry saturation). The T$_{90}$ duration
of the GRB in the 15-500 keV band is 124 s.

Since the GRB peaks during the satellite slew, we could not use the
instrumental coordinates to extract the peak spectrum.
Therefore we made an image selecting a time interval of 1 s around the GRB main peak
(t = 54 s in Fig. \ref{lc}). The high count rate at the peak allowed
us to firmly establish the detector coordinates of the GRB even with this
short integration time and thus to extract its peak
spectrum. Since  IBIS/ISGRI is a coded mask imaging instrument,
the background can be estimated simultaneously with the
source flux, using the Pixel Illumination Function (PIF; \cite{skinner}).
The spectra have been
extracted computing one PIF for each energy bin (128 linearly spaced
bins have been used between 19 keV and 1 MeV). Since a fully calibrated
spectral response matrix for sources at large off-axis angles is not yet available,
we divided the count spectrum by the closest (in detector coordinates)
count spectrum of the Crab Nebula. The resulting photon
spectrum can be well fitted by a   power law model
($\chi^2$/dof = 8.66/8) with photon index $\Gamma$ = 1.73$_{-0.17}^{+0.16}$
(90\% confidence level). The flux is $\sim$1.9 photons
($\sim$ 1.7$\times$10$^{-7}$ erg) cm$^{-2}$ s$^{-1}$ ($\sim$6.5 crab)
in the 20-200 keV energy range.

\begin{figure}
      \hspace{0cm}\psfig{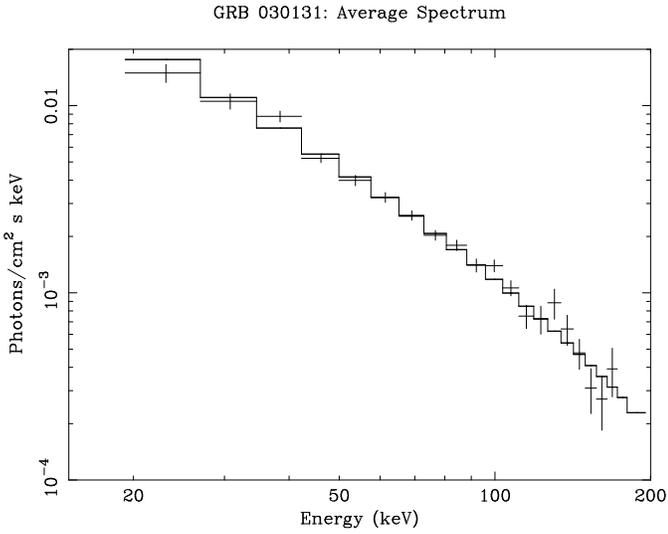}
      \caption[]{Time averaged IBIS/ISGRI spectrum of GRB 030131. Data and best fit model are shown.}
         \label{total}
   \end{figure}

To properly extract the total GRB flux and spectrum, we derived
the GRB detector coordinates at various time intervals
for the entire duration of the event.
The first interval, corresponding to the stable pointing, lasts 20 s.  The
following 30 intervals last 3 s each and, finally, for the faint tail of the burst,
four intervals with durations of 5, 5, 10 and 20 s were used.
\begin{figure}
      \hspace{0cm}\psfig{figure=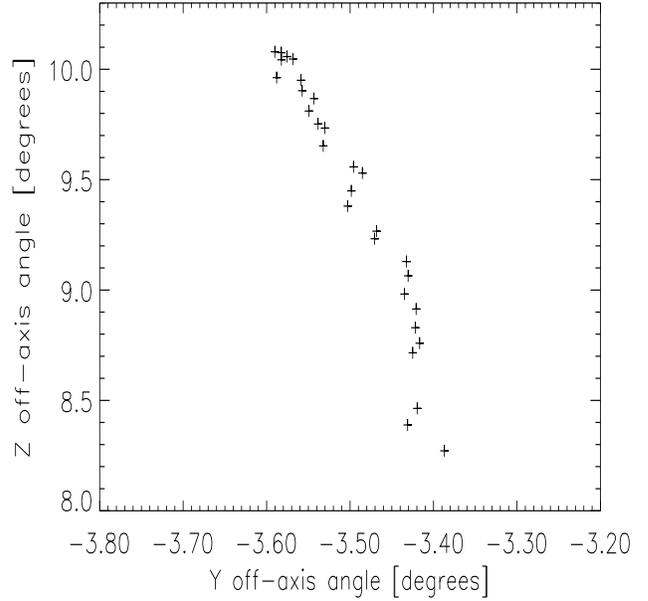,width=8.8cm,height=9.cm}
      \caption[]{Positions of GRB 030131 in detector coordinates
as a function of time. Note the different scales of the two axes.}
         \label{positions}
   \end{figure}
The coordinates as a function of time are shown in
Fig. \ref{positions}, where one can see the apparent drift of $\sim$2$^{\circ}$
of the GRB in the IBIS field of view
during the slew. To obtain the total (time averaged) spectrum,
the  35 spectra  extracted for the time intervals,
corresponding to a net integration time of 147.2 s, were co-added.
We derived the photon spectrum (shown in  Fig. \ref{total}) in the same way
that we did for the peak spectrum. The GRB is clearly detected up to 200 keV.
In this case a single power law
model does not provide a satisfactory fit ($\chi^2$/dof = 39.84/18), indicating that a
model which includes a spectral break would give a better
representation of the data. A fit using the Band model
(\cite{band}) yields a break energy $E_{0}$=70$\pm$20 keV, a low-energy
power law index $\alpha$ = 1.4$\pm$0.2 and a
high-energy photon index $\beta$=3.0$\pm$1.0 ($\chi^2$/dof = 22.54/16).
The fluence in the 20-200 keV band is $\sim$73.8 photons cm$^{-2}$ (7 $\times$
10$^{-6}$ erg cm$^{-2}$). To derive this value we have assumed that during the 4 central
telemetry gaps the GRB had its average spectrum and intensity.
Extrapolating the spectrum to the BATSE energy
range (50-300 keV) we obtain a fluence of $\sim$23 photons cm$^{-2}$ (3.2 $\times$
10$^{-6}$ erg cm$^{-2}$).
The values of peak flux and fluence derived above are consistent
to within a factor of 2 with the ones measured with Ulysses. This indicates
that our method does not suffer from large systematic errors and that the
short telemetry gaps do not influence these values significantly.

We have also investigated the spectral evolution.
As in the case of the peak spectrum, the data for the individual time
intervals can be fitted with a
single power law, without evidence for a spectral
break. The photon index as a function of time is plotted
in Fig. \ref{lc}. A clear hard-to-soft evolution is seen.
A hardening trend can also be seen
corresponding to the rise of the second and third (main) peak
 with a softer spectrum for the latter peak.
This correlation between light curve peaks and spectral hardening
has already been reported in other bursts (e.g. \cite{ford}).

\section{Discussion}

The IBIS/ISGRI time-resolved spectroscopy
of GRB 030131 is consistent with the overall hard-to-soft evolution
observed with BATSE in many  brighter GRBs, for which
this kind of analysis was possible (e.g. \cite{preece}, \cite{ford}).
The fluence of GRB 030131 is an order of magnitude smaller than those of the
bursts studied by those authors, indicating that such spectral behaviour applies also
to fainter GRBs. Clear evidence of this was also reported
in the GRBs studied with {\it Beppo}SAX in the 2-700 keV energy range (e.g. \cite{ff}, \cite{ff2}).
While BATSE could better constrain the break energy and the high-energy slope
of the Band function, thanks to its higher relative sensitivity above 200 keV,
IBIS/ISGRI allows more
detailed studies of the low-energy part of the spectrum for relatively low fluence GRBs.

In the framework of
the internal fireball shock model (\cite{rees94}), and in particular of the
Synchrotron Shock Model (\cite{tavani}), the hard-to-soft evolution can be interpreted in
two ways.  The first possibility is a decrease
of the magnetic field in the postshock region as a consequence
of the postshock flow expansion; the second is a postshock decrease of
the index of the particle distribution function as a consequence of strong
cooling processes affecting the particle energy distribution for
dynamical flow times larger than the radiating timescale.
The two effects are not distinguishable
in our case, since the time-resolved spectra do not
have enough statistics to constrain a spectral break and hence a low-energy
and a high-energy spectral index. The soft-to-hard evolution observed during
the rise of the individual peaks, on the other hand, can be caused by an increase of the local
magnetic field at the shock. Several authors reported
that the duration of single pulses in GRB time histories
is energy dependent (e.g. \cite{link}), with longer durations at lower energies,
resulting
in a hardening of the spectra before the peaks and a softening afterwards.

The optical transient associated with GRB 030131 indicates that
we can classify it as an ``optically dim'' GRB.
In fact it is as faint as (or even
fainter than) the transient associated with GRB 030227, which is the only other
INTEGRAL GRB with a firmly
established optical conterpart, and was detected at R$\sim$23.3
12 hours after the burst (\cite{mere3}). It is also
comparable to GRB 021211. This event is also considered an
optically dim burst since it
was detected at R$\sim$18.2 1.3 hours after the prompt emission
(\cite{fox2}) and was fainter than  R$\sim$22.5 after
12 hours (\cite{klose}). In addition GRB 030131 is located at much higher Galactic
latitude (b $\simeq$ 81$^{\circ}$) which implies smaller foreground optical extinction.
This indicates that, despite the efforts of observers, in
some cases, optical follow-up with small telescopes is not an easy task,
even less than 1 day after the burst (e.g \cite{fynbo}).
The prompt localization of
GRBs is hence a high priority in order to achieve a successful follow-up.
The results on GRB 030501 (\cite{beckmann}) (an alert 30 s after the start of the
GRB with an uncertainty of 4.4$'$; IBAS Alert 596) show that
IBAS is now able to provide this service.

\begin{acknowledgements}
This research has been supported by the Italian Space Agency.
KH is grateful for Ulysses support under JPL contract 958056,
and for support of the IPN under NASA grant NAG5-12614. DHH
acknowledges support by NASA. RH acknowledges
the support by Prodex Project 14527.
\end{acknowledgements}

\end{document}